\documentclass[preprint,journal=jacsat]{achemso}
\usepackage{graphicx} % Required for inserting images
\usepackage{hyperref}
\usepackage{lineno}  % line numbers
\usepackage{xcolor}
\usepackage{mhchem}
\usepackage{natmove}
\usepackage{braket}
\usepackage{cleveref}
\usepackage{epstopdf}

\AppendGraphicsExtensions{.tif}

\colorlet{review}{red}

\title{Continuous Local Symmetry: Connection to Reactivity and Recognition}

\author{Duc Anh Lai}
\author{Devin A. Matthews}
\email{damatthews@smu.edu}
\affiliation{Department of Chemistry, Southern Methodist University, Dallas, TX 75275, USA}

\begin{document}

\begin{tocentry}
\includegraphics[width=8cm]{TOC1.tif}
\end{tocentry}

%\linenumbers
\begin{abstract}
Symmetry is one of the most beautiful yet mysterious concepts in science. In chemical systems, presence of local symmetries at specific fragments often serve as driving forces behind many physicochemical properties, including stability, spectroscopy, and reactivity. Moreover, degree of symmetry varies continuously with molecular dynamics and intermolecular interactions, making it a hidden but decisive factor. In this study, we propose a theoretical framework to quantify continuous degrees of symmetry and chirality localized within constrained regions of a molecular environment. Application of this method to reaction sites of dendralene molecules reveals strong correlations between local symmetry and molecular stability, parity-dependent behavior, and Diels--Alder reactivity. Additionally, representations of local chirality (chirotopicity) fields in porphyrins uncover unique signatures accounting for the chirality recognition power. Overall, these findings highlight the potential of local symmetries within a molecular framework on predicting chemical properties.
\end{abstract}

\maketitle

\section{Introduction}

Since the dawn of chemistry and physics, symmetry has served as both a conceptual property and a computational tool, as represented in a wide range of fundamental discoveries.
%\textcolor{blue}{[do we actually need to cite some?]}
In chemistry, the symmetry of nuclear and electronic distributions forms the foundation for numerous theories, such as valence shell electron pair repulsion (VSEPR) theory, frontier molecular orbital (FMO) theory, selection rules in spectroscopy, quantum degeneracy, and computational methods for structure and potential energy surface prediction. In practice, group theory and symmetry analysis provides an excellent explanation for many chemical behaviors in spectroscopy, chemical reactions \cite{woodward_conservation_1969, pearson_symmetry_1986}, catalysis \cite{caldow_symmetry_1970}, chemical reactions \cite{maksic_symmetry_1986, hoffmann_interaction_1971}, molecular recognition \cite{cram_host-guest_1974}, self-assembly \cite{sang_symmetry_2019, sang_hierarchical_2022}, and many others. Despite the importance of symmetry in chemical contexts, traditional analysis only characterizes molecular symmetry as an absolute property, i.e. whether a molecule possesses a certain symmetry element or not. Beyond the static equilibrium picture, the symmetry of molecular systems can also be considered from a dynamic or trajectory viewpoint. For example, rotational isomerism along the \ce{C-C} $\sigma$ bond in ethane causes a continuous change from $D_{3d}$ to $D_{3h}$ symmetry. Also, small distortions due to Jahn-Teller effects, thermal dynamics, solvent fluctuation, or crystal packing often deform ideally high symmetry, or inversely, thermally surmountable barriers can yield average structures with higher symmetry. As a result, many dynamic molecules are best understood as time-averaged symmetric ensembles. Spontaneous distortions taking place in organometallic complexes, piezoelectric materials, and crystallographic systems introduce symmetry variations and modify optical, electric, and magnetic properties \cite{bersuker_jahn-teller_2017, bersuker_jahnteller_2021, shi_symmetry_2016}. Organic chemistry is perhaps the area that most utilizes symmetry analysis of molecular systems. In pericyclic reactions, Woodward and Hoffmann's rules state that the symmetry of interacting molecular orbitals must be constantly conserved for the reaction to proceed under thermal conditions, otherwise photoexcitation is necessary \cite{woodward_conservation_1969}. Moreover, biological macromolecules appear to exhibit cyclic ($C_n$), dihedral ($D_n$), and cubic ($I, O, T$) symmetries, governing a wide variety of functions \cite{duarte_exploring_2022}. However, their conformations may undergo drastic changes upon the binding of modulators \cite{keinan_quantitative_2000}. Therefore, a concept of continuous molecular symmetry is essential and advantageous. 

The first attempt to quantify continuous symmetry in chemistry originated from Guye's asymmetry product which empirically correlated with the optical activity \cite{petitjean_chirality_2003}. Subsequently, a number of symmetry measures based on molecular geometry \cite{zabrodsky_continuous_1992}, volume overlap \cite{mezey_degree_1991}, molecular surface topology \cite{ferrarini_assessment_1998}, and wavefunction or electron density \cite{grimme_continuous_1998, aucar_relationship_2024}, were introduced. Numerous studies have successfully applied these approaches to rationalize experimental observation \cite{shah_calochorturils_2025, guo_switchable_2022}. Since electron configuration dictates properties of a molecule, electronic density- or wavefunction-based methods are perhaps more suitable than structural and topological methods when one focuses on chemical behaviors and transformations. Particularly, electronic properties of chemical substituents may have significant impact on a local scaffold in resonance structures and charge transfer systems. Especially, when a molecule is influenced by surrounding molecules, or external perturbations, such as electric and magnetic fields, the electron density may not coincide with nuclear configuration. Moreover, a phase transition may occur in ultrafast dynamics, e.g. femtosecond to attosecond timescale, without a reorganization in nuclear positions, resulting in a spontaneous electronic symmetry breaking. It is also suggested that geometry alone may not suffice, other aspects, such as electronic effects, must also be considered when evaluating chirality \cite{fowler_vocabulary_1992}, and correlating with functional predictions \cite{grieder_relation_2025}. Among the aforementioned measures, Grimme's method is the current choice for calculating global symmetry from molecular electronic structure calculations \cite{guo_switchable_2022}. 

In the modern context, and particularly as chemical structures of interest grow more complex with new experimental and computational techniques, the use of symmetry considerations for the entire molecule is restricted as perfect, global symmetry elements are no longer available for large and flexible systems. This leads to the adage, heard at times by the authors, that ``all interesting molecules are $C_1$.'' However, in many situations a non-symmetric molecule may possess local segments that are fully or nearly symmetric, and these local symmetries can play a dominant role in representation of molecular properties such as spectroscopic shifts, bond strength, stability and reactivity. In particular, qualitative assessments of local symmetry are usually invoked to identify equivalent atoms, which determines distinct chemical shifts in nuclear magnetic resonance and other spectroscopies. Additionally, even non-symmetric molecules can inherit the highest symmetry from its symmetrical substructures, e.g., propylene can be reduced to $C_s$ semi-symmetry if one considers a chemical process that solely involves the $\pi$ orbitals. Furthermore, global and local symmetry are likely dissimilar in transition metal systems due to bulky ligands. According to ligand field theory, orbital splitting in such molecules is more dependent on microscopic environment around the central metal than on the global symmetry group of the entire complex. In many cases, coordination complexes take over pseudo-symmetry of the local symmetry at the metal site. On the other hand, defects in solid-state materials break down local symmetry and profoundly alter electrical conductivity, optical activity, hardness and ductility, ferromagnetism, catalytic selectivity, and activity. 

The idea of local symmetry in chemistry was first discussed in the perspective of Mislow and Siegel, connecting the local symmetry of segmented fragments to the context of molecular environment \cite{mislow_stereoisomerism_1984}. The authors proceed from the viewpoint that local symmetry generally exists at any point within the molecular framework, regardless if such point is an atomic center or not. From that starting point, they then develop the concept of the ``chirotopicity field'' as a local analogue of chirality, which exists even in overall achiral molecules. Combining this viewpoint with the concept of continuous symmetry, this notion implies that local symmetry should vary smoothly throughout the molecular environment, forming a continuous field that covers the entire molecule and its surroundings. Currently, there are few published techniques to calculate the local symmetry. One approach is to partition the molecular structure into fragments \cite{lipinski_local_2014}. The main disadvantage of this method is that structural and electronic effects from the surrounding moeities are ignored. With a similar idea, Avnir and co-workers divided coordinate complexes into spherical shells, and compute the perturbative effects of outer shells on the degree of local chirality at the center metallic atom \cite{alvarez_continuous_2005}. However, the method is largely biased by the choice of reference structure. As a consequent, the results are less selective when comparing across different systems. Grimme\cite{grimme_continuous_1998} suggested that his method of computing continuous symmetry could also be partitioned into contributions from localized occupied orbitals. However, this approach does not address symmetry at arbitrary points and is inherently tied to the local topology of the orbitals.

In this manuscript, we propose a novel method to quantify the continuous local symmetry of a chemical system from the total electronic density. We first demonstrate how local symmetry correlates with chemical stability and reactivity of asymmetric diene units in dendralenes. Subsequently, we study the extent of local chirality of substituted porphyrins, and how that the local chiral information presents key signatures for the chiral recognition strength of porphyrins toward chiral carboxylic acids. These preliminary applications suggest a deep connection between continuous local symmetry and observable molecular properties. 

\section{Theoretical Methods}

In this work, the degree of local symmetry with respect to a symmetry element $R$ is calculated as,
\begin{equation}
    S{(R)}=1-\frac{||D_R-D_0||_F}{||D_R||_F+||D_0||_F}
\label{eq:S}
\end{equation}
where $D_0$ is the molecular relaxed one-electron density matrix (1RDM) projected onto a local, isotropic basis centered at a chosen point in space. The image density $D_R$ is then obtained by transforming $D_0$ according to the matrix representation of the symmetry operation. The subscript $F$ denotes the Frobenius norm. This measure provides a value between zero (no symmetry, i.e. the original and image densities do not overlap) to one (exact symmetry). Additionally, by leveraging the electronic density directly, densities from any ab initio or semi-empirical computational method may be employed, as well as densities tabulated in real space.

While pre-chosen symmetry elements may be used in the computation of the symmetry measure, more commonly the parameters defining the element (e.g. axis of rotation or normal vector of a reflection) may be optimized such that $S(R)$ is maximized, although it is required that the symmetry axis pass through the chosen projection point. In many cases, the optimization surface of the symmetry axis Euler angles permits multiple local minima---a feature which merits further investigation in combination with the corresponding local axis frames.

As chirality emerges when a molecule lacks second-order mirror symmetry, we can compute local chirality (chirotopicity) by exhaustively finding optimal improper rotations for the local structure. Thus, local chirality can be obtained from the maximum value of local symmetry with respect to either reflection ($\sigma=S_1$), inversion ($i=S_2$) or higher improper rotation ($S_3, \ldots $) as,
\begin{equation}
    C=1-\max\{S(\sigma), S(i),S(S_3),\ldots\}
\end{equation}
where each of the symmetry measures is optimized with respect to the orientation of the symmetry element. Obviously, when there exists an ideal improper rotation, $C = 0$. On the other hand, any imperfect symmetry gives rise to a non-trivial value of $C$. At the upper limit $C=1$, the second-order mirror symmetry vanishes and the molecule (or molecular region) is ``maximally chiral''. 
%Because local symmetry is normalized (eqn \ref{eq:S}), $C$ also spans the domain [0, 1], where 0 means chiral, and 1 means achiral. 

% As previously noted\cite{grimme_continuous_1998}, $S_3$ axes and above are rarely the elements with highest measure, and computing $S(\sigma)$ and $S(i)$ only suffices.
%Even if a higher $S_n$ exists, there is likely lower $S_n$ axis in the system. Physically, the highest common improper rotation axis observed in highly symmetric structures is usually $S_6$ or $S_8$. Higher Sn axes appear in particular systems, such as $C_{60}$ fullerene, ferrocene, and other engineered materials. 

\section{Results}
\subsection{Local symmetry: A predictor of dendralene reactivity and stability}

As noted earlier, the local symmetry of a molecule may diverge significantly from its global symmetry. This distinction is particularly critical because most chemical processes occur locally at specific functional sites within a molecule. Therefore, understanding local symmetry deviations is important, yet underexplored, for predicting reactivity, selectivity, and other chemical properties. In this context, the family of dendralenes stands out as an illustrative example: these molecules possess regional symmetries that are not immediately apparent from their entire structure, making them ideal systems for investigating how local symmetry variations influence chemical functionalization patterns.

Dendralenes are a class of branched, acyclic oligo-olefinic hydrocarbons, together with linear polyenes, annulenes, and radialenes. Although the unbranched systems, e.g., linear polyenes and annulenes, have been comprehensively studied since the 1970s, the two branched classes, including [$n$]dendralenes, have just received significant attention recently \cite{mackay_demystifying_2013}. Thanks to the groundbreaking studies of Hopf, Fallis, and Sherburn, to name a few, [3--12]dendralenes and their substituted derivatives have been successfully synthesized and investigated in terms of chemical stability, Diels--Alder (DA) reactivity, and spectroscopic properties \cite{hopf_dendralenesneglected_1984,woo_indium-mediated_1999,fielder_first_2000,george_general_2019,saglam_discovery_2016}. It is evident that dendralenes are beneficial molecules in organic synthesis, such as step-economic target synthesis on account of their ability to participate in diene-transmissive Diels--Alder reactions \cite{saglam_synthesis_2016,pronin_synthesis_2012}. 

\begin{figure}
    \centering
    \includegraphics[width=\linewidth]{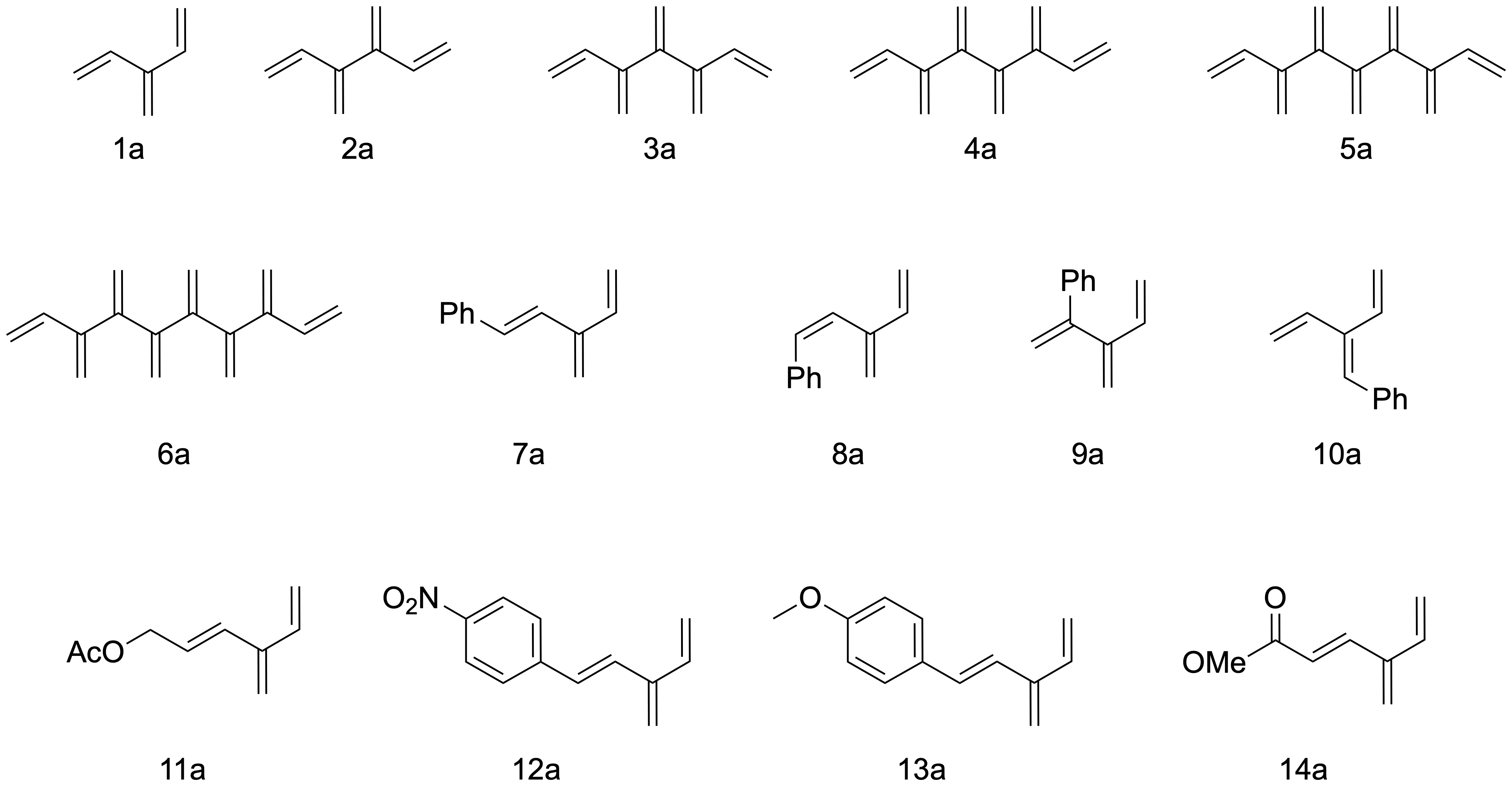}
    \caption{[3--8]Dendralenes and derivatives.}
    \label{fig:den_mols}
\end{figure}

A single diene, e.g. 1,3-butadiene, can adopt \emph{s-cis}, \emph{s-trans}, and gauche conformations depending on the dihedral angle around the internal $\sigma$ bond. Since [$n$]dendralenes possess $n-1$ diene units arranged in a branched configuration, $sp^2$ carbons are twisted to $sp^3$-like hybridization, mostly due to the steric hindrance of inner hydrogens. As a consequence, the torsion angles in [$n$]dendralenes are varied largely, featuring several abnormal physiochemical properties. Efforts were made to analyze [3--8]dendralenes' conformers, indicating that imperfect symmetries of odd and even dendralenes represent the parity-dependent behavior, the so-called alternation feature, represented in UV-Vis, \ce{^1H} and \ce{^{13}C} NMR spectra, and mono-cycloaddition DA reactivity \cite{saglam_discovery_2016}. The alternation behaviors are genuinely results of local diene unit properties. For example, species from [4]dendralene to [12]dendralene exhibit a maximum UV-visible absorption at approximately the peak of 1,3-butadiene, suggesting that the conjugation of these systems is restricted to locally single diene structures \cite{payne_practical_2009}. It is also clear that the \emph{s-cis} diene units, which are necessary for cycloadditions, are no longer coplanar, leading to unpredicted reactivity in the DA pathway \cite{saglam_discovery_2016}. In this respect, breaking of the perfect $C_{2v}$ symmetry from an ideal \emph{s-cis} diene may regulate the cycloaddition reaction of the dendralenes. We hypothesize that the partial symmetry descent in local quasi-\emph{s-cis} dienes is associated with the alternation behavior of the [$n$]dendralenes.  

\begin{figure}
    \centering
    \includegraphics[width=\linewidth]{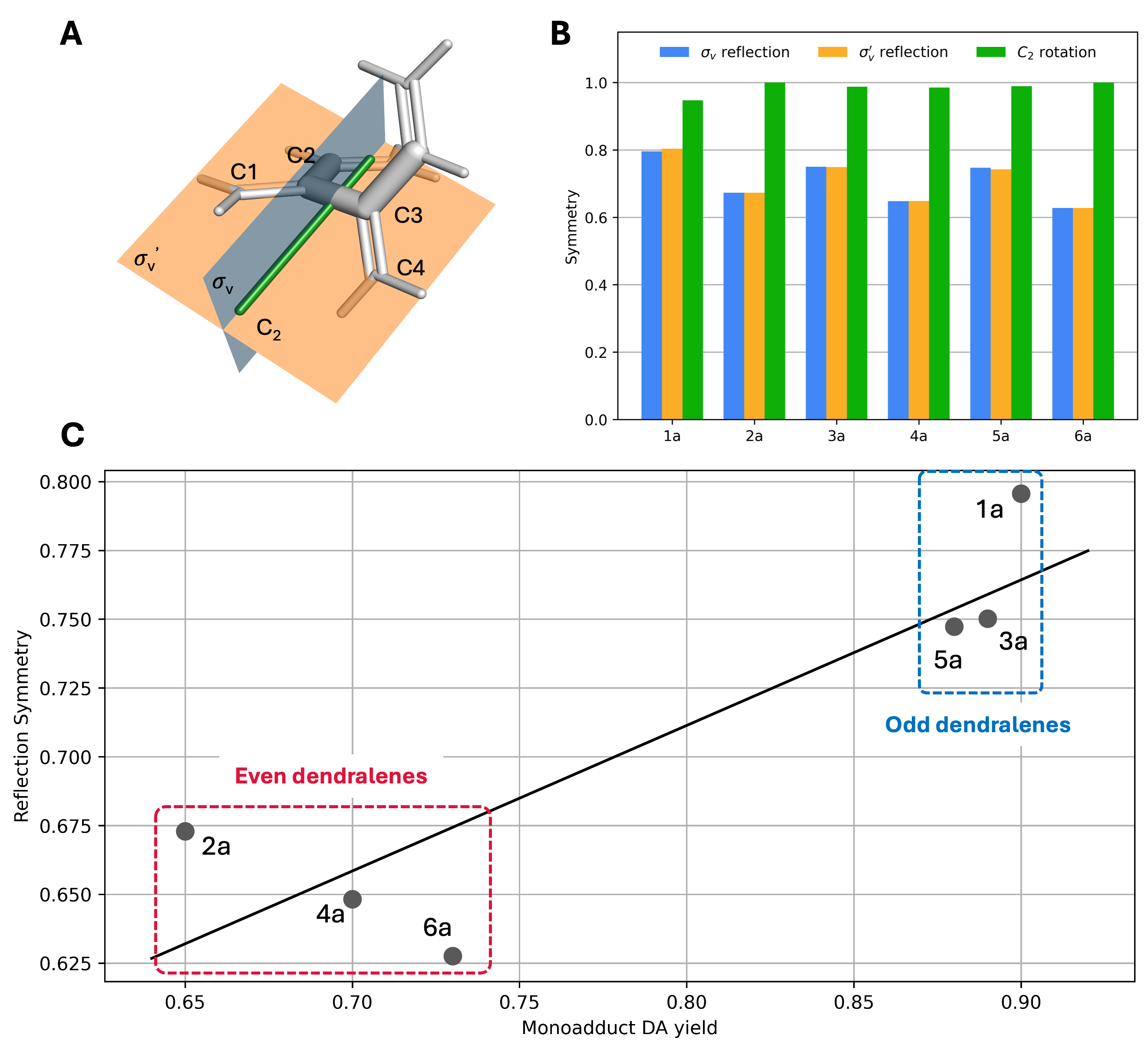}
    \caption{\textbf{a)} Local quasi-$C_{2v}$ symmetry elements in [4]dendralene. \textbf{b)} Local symmetry measures of [3--8]dendralenes. \textbf{c)} Correlation of local reflection symmetry ($\sigma_v$) with monoadduct Diels--Alder reactivity. Linear regression (solid line) yields a coefficient of determination of $R^2 = 0.77$.}
    \label{fig:den_alternation}
\end{figure}

We performed a local symmetry analysis on quasi-cis dienes in [3--8]dendralenes (\textbf{1}--\textbf{6a} in \Cref{fig:den_mols}) with respect to $C_{2v}$ symmetry. 
% The most stable conformer of each molecule was retrieved from Saglam's work \cite{saglam_discovery_2016}. First, we re-optimized molecular geometries using DFT/B3LYP/6-31G(d) with D3 dispersion correction and chloroform SMD solvation model \cite{marenich_universal_2009}. 
Local $C_{2v}$ symmetry of a C1--C2--C3--C4 diene consists of 3 non-trivial symmetry elements: a two-fold rotation ($C_2$) and two vertical reflection planes ($\sigma_v$ and $\sigma_v'$) as depicted in \Cref{fig:den_alternation}\textbf{a}. The $C_2$ axis was defined by a vector connecting the centers of the C1--C4 and C2--C3 vectors. The $\sigma_v$ plane bisects the diene fragment with its normal parallel to the C1--C4 vector. The other mirror plane, $\sigma_v'$, was determined such that its normal was obtained from the cross product of the $\sigma_v$ normal and the $C_2$ axis. By construction, the two reflection planes are mutually perpendicular and both contain the rotation axis. All local symmetry computations were evaluated with at the midpoint of the C1--C4 distance.

As shown in \Cref{fig:den_alternation}\textbf{b}, local symmetry measures clearly show the alternation feature between even and odd members of the dendralene series. We found that the symmetry measures for the two reflection planes are nearly equivalent, while local dienes exhibit an almost perfect $C_2$ symmetry. Therefore, the \emph{s-cis} $C_{2v}$ symmetry is deformed to a quasi $C_2$ symmetry in twisted dienes primarily by torsion of the C1--C2 and C3--C4 bonds. The lowest value of S($C_2$) is 0.95 in [3]dendralene, and full $C_2$ symmetry appears in local dienes of [4]dendralene and [8]dendralene. The [4] and [8]dendralene conformers also adopt $C_2$ point group for the total symmetry, showing that a connection still exists between some local symmetries and the global point group. However, as discussed below, it is the reflection and not rotation symmetry which determines local reactivity. While not required by the molecular point group, the two reflection symmetries in [4] and [8]dendralenes show no difference in magnitude. In contrast, among [3] and [5--7]dendralenes, the lower S($C_2$) measure corresponds to a (slightly) larger difference between the two mirror symmetry measures. This is reasonable as deviations from $C_2$ indicate additional perturbations which necessarily affect the C1--C2 and C3--C4 local geometries and electronic structures differentially.

More importantly, a strong correlation was observed between local mirror symmetry characteristics and mono-adduct DA reactivity. In a concerted DA pathway, two new $\sigma$ bonds are formed synchronously at the termini of the diene. As this process is governed by constructive orbital overlap between the HOMO of the diene and the LUMO of the dienophile (or vice versa), orbital symmetry considerations play a decisive role, and the planar \emph{s-cis} conformation of the diene is preferred. The gauche-like functional site in dendralenes, however, make their DA reaction outcomes difficult to generalize. Experimental studies showed that odd dendralenes have substantially higher reactivity toward cycloaddition pathways than their even neighbors \cite{paddon-row_origin_2011, payne_practical_2009}. This observation can be rationalized in terms of the local symmetry of the bisector plane $\sigma_v$ in the local diene fragment. The highly reactive odd dendralenes exhibit significantly higher reflection symmetries compared to the nonreactive even dendralenes (\Cref{fig:den_alternation}\textbf{c}).  In the specific case of [$n$]dendralenes, the diene distortion consists primarily of torsion about the central \ce{C-C} bond. The torsional angle itself does also roughly correlate with mono-adduct reactivity and the resultant alternation, although not as significantly as the local reflection symmetry measure (Figure~S1). However, the local symmetry analysis is vastly more versatile as it can describe a wide range of local internal coordinate distortions, as well as complex mixtures thereof.

\begin{figure}
    \centering
    \includegraphics[width=\linewidth]{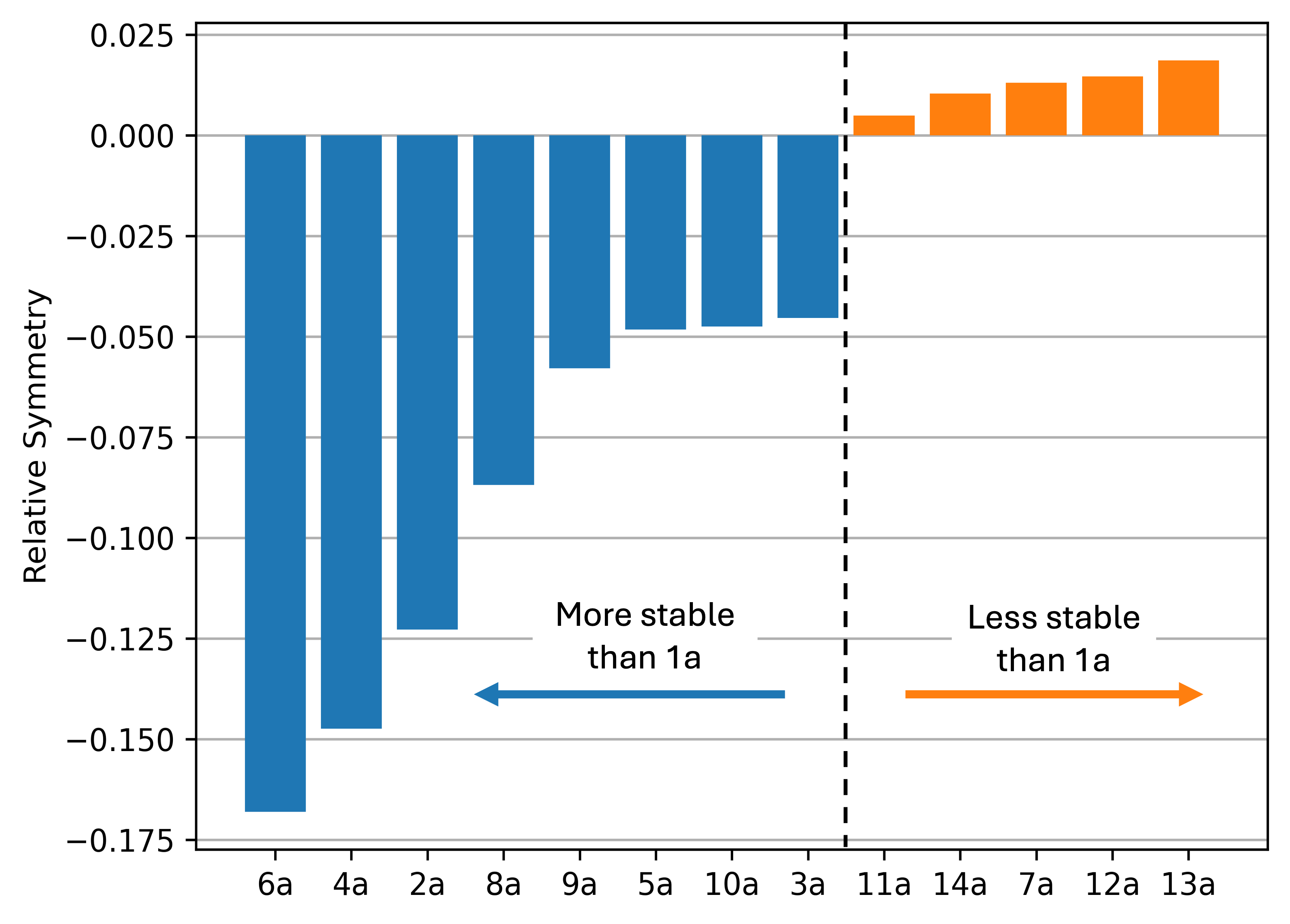}
    \caption{Difference in local reflection symmetry ($\sigma_v$) of substituted dendralenes with respect to [3]dendralene (\textbf{1a}). Dendralenes with lower reflection symmetry are more stable than \textbf{1a} (blue), while those with higher reflection symmetry are less stable (orange). Note that species are arranged in order of local symmetry.}
    \label{fig:den_stab}
\end{figure}

The symmetry distortion of local diene groups in unsubstituted dendralenes is mostly due to the conformational steric effect of cross-conjugation. Subsequently, we included other derivatives into our test set to see whether the proposed local symmetry analysis captures electronic perturbations from chemical substituents to dendralenes in addition to purely steric effects. The substitutions vary in type of functional groups and position of modification (\textbf{7}--\textbf{14a} in \Cref{fig:den_mols}). Our next analysis focuses on the stability of these dendralene derivatives. The DA dimerization mechanism is proposed to explain dendralenes' stability in pure solution. In this regards, two dendralene molecules can react with each other through a closed-shell singlet bis-pericyclic transition state \cite{toombs-ruane_dielsalder_2012, paddon-row_origin_2011}. Since the essence of this mechanism is associated to the DA mechanism, it turns out that the plane of symmetry can also classify the stability of dendralenes. As shown in \Cref{fig:den_stab}, all molecules that are more stable than [3]dendralene (\textbf{1a}) have lower mirror symmetry values. The even dendralenes reported to be least reactive among dendralene family, displaying negligible decomposition over four weeks at ambient temperature, present the highest degree of distortion in terms of mirror symmetry. Among odd-membered unsubstituted dendralenes, increasing the number of double bonds results in more stable molecules, which is represented by a consistently decreasing trend in reflection symmetry. Theoretically, 1Z-, 2-, and 3'-phenyl-[3]dendralenes destabilize the biradicaloid transition structure, indicated by extended lifetime at room temperature \cite{toombs-ruane_dielsalder_2012, saglam_synthesis_2016}. The bisector plane of symmetry correctly predicts that these three mono-substituted [3]dendralenes are less symmetric than the unsubstituted counterpart. In contrast, other substituents at C1 of [3]dendralene, including 1E-phenyl, 4-nitrophenyl, 4-methoxyphenyl, methoxycarbonyl, and acetoxymethyl exhibit facile DA dimerization behavior and are predicted to be less symmetric with respect to the $\sigma_v$ plane than [3]dendralene. As the local symmetry measure utilizes the electronic density, it is sensitive to the resonance effects of $\pi$-conjugation and induction from electron withdrawing groups. Purely geometric measures such as diene dihedral angle capture the gross trend, but are not sensitive to subtle electronic features, for example mispredicting \textbf{11a} as more stable than \textbf{1a} (Figure~S2).

%\textcolor{blue}{TODO}
%This result complements with the correlation between mirror symmetry in dendralenes and cycloaddition reactions. 

\subsection{Chiral microenvironment: A hidden factor enabling chiral recognition phenomena}

Chirality has always been a central subject in a wide range of chemical fields such as medicinal chemistry, catalysis, and materials science \cite{sallembien_possible_2022, pavlov_hiral_2019, takahashi_origin_2019, nguyen_chiral_2006}. Although two stereoisomers of a chiral molecule share identical connectivity, they interact completely differently with other chiral environments, such as catalysts, enzymes, receptors, or polarized light, resulting in complex chemical behavior \cite{berthod_chiral_2010}. This subtle asymmetry recognition governs many chemical phenomena, from enantioselective synthesis to therapeutic and toxic effects of drugs \cite{cui_bionic_2024, daskova_turning_2022, che_hierarchical_2025}. Moreover, the spontaneous induction, amplification, and transfer of chirality provide deep insights into reaction mechanisms, stereoselectivity, and even the origin of biological homochirality \cite{mayer_nonlinear_2022, sallembien_possible_2022, buhse_spontaneous_2021}.

Local symmetry directly connects to the concept of local chirality. By definition, chiral molecules lack second-order symmetry elements (improper rotations), typically a reflection plane or inversion center. Consequently, chirality is an intrinsic property of a molecule whose non-superposable mirror images form distinct enantiomers. On the other hand, a molecule is categorized as achiral if it is superposable on its reflection (inversion, etc.). By quantifying the degree of local symmetry, we can  quantify chirality, discretized in a constrained region, which we term the chiral microenvironment. In this section, we present how the local chiral microenvironment manifests chiral recognition phenomena in substituted (5,10,15,20)-tetraphenyl-octamethylporphyrins. 

\begin{figure}
    \centering
    \includegraphics[width=0.5\linewidth]{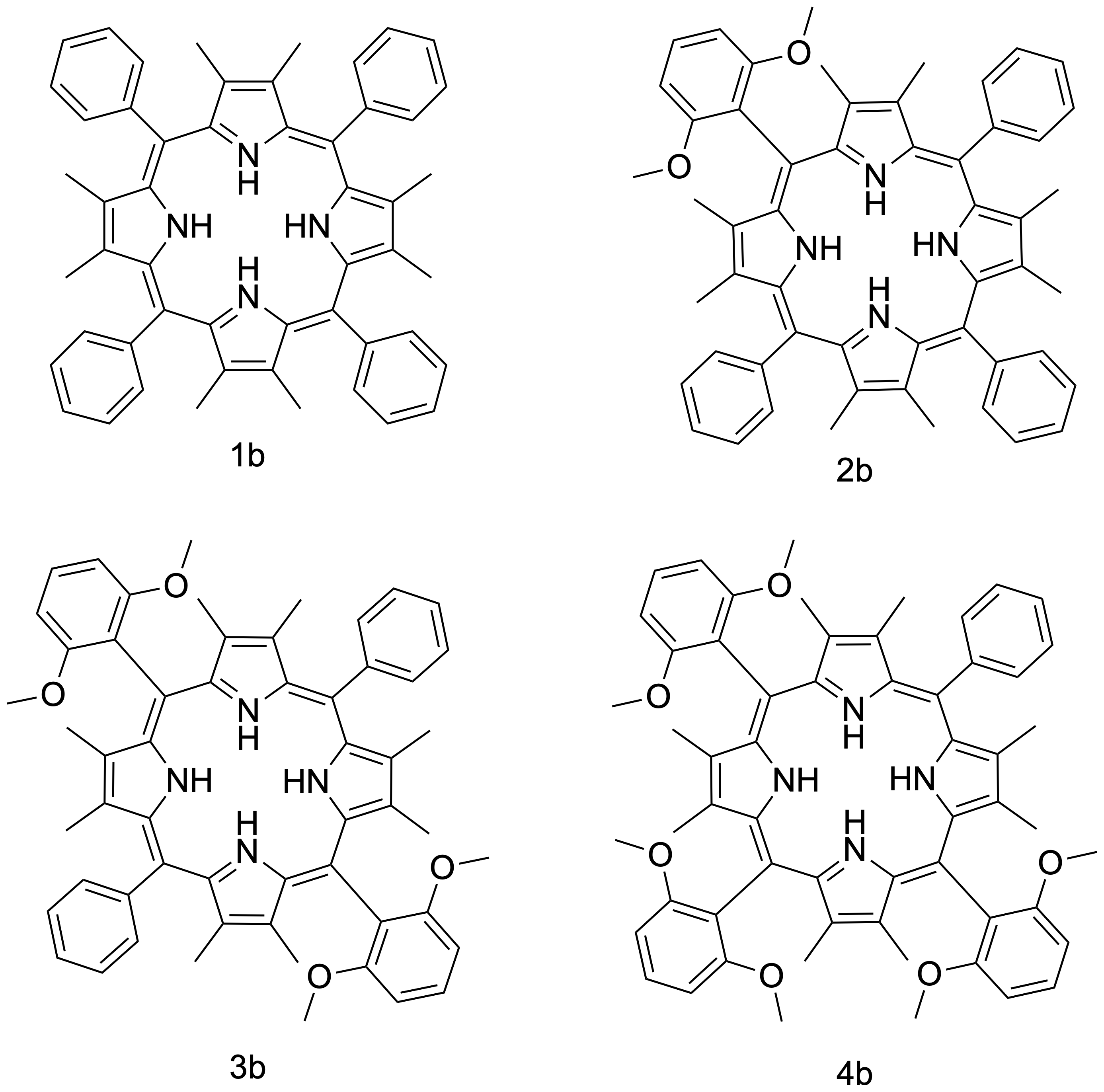}
    \caption{Tetraphenyloctamethylporphyrin and dimethoxyphenyl-substituted derivatives. The diprotonated structures are depicted in accordance with the carboxylic acid-complexed crystal structure.}
    \label{fig:por_mols}
\end{figure}

Porphyrins have been extensively studied for many decades from pure chemistry and biology to applied studies in the fields of pharmaceutical science, materials, devices, and many others \cite{hirotoSynthesisFunctionalizationPorphyrins2017}. In chemical aspect, porphyrin structures possess a macrocyclic skeleton consisting of four pyrrole rings bridged by \emph{meso} methines in 1,3-positions, forming one of the most highly symmetrical molecular shapes in macromolecular chemistry. The most interesting feature of porphyrins is their flexibility to acquire distinctive chemical behavior through engineered functionalization \cite{hirotoSynthesisFunctionalizationPorphyrins2017}. Obviously, there are 12 available positions on the porphyrin core that can be modified: four \emph{meso} positions on the bridges and eight $\beta$ positions on the pyrrole units. While $\beta$ substitutions are widely considered to be the most powerful approach to alter a porphyrin's properties and reactivity, \emph{meso} phenyl substitutions have been shown to disrupt the overall planarity of the system, breaking the nominal $D_{4h}$ symmetry, and even producing chiral derivatives.\cite{furusho_chirality-memory_1997,mizuno_chirality-memory_2000,syrbu_phenyl-substituted_2004} In this manuscript, we focus on a class of porphyrins with methyl groups in all $\beta$ positions, and either phenyl or 2,6-dimethoxyphenyl substitutions in \emph{meso} positions (\Cref{fig:por_mols}). \textbf{1b} is the ``base'' tetraphenylporphyrin, whereas \textbf{2b}--\textbf{4b} successively substitute dimethoxyphenyl groups, producing a range of global symmetries and, as discussed below, local chiral environments. As \textbf{1b} exhibits the highest $D_{2d}$ symmetry, it is achiral. In contrast, \textbf{2b} and \textbf{4b} possess a single $C_2$ rotation, while \textbf{3b} presents $D_2$ symmetry with three rotation axes. All substituted porphyrins are therefore chiral. Due to their partial symmetry, the two sides of the molecular plane of all systems behave identically.

These particular porphyrins are reported as effective hosts for sensing asymmetric carboxylic acids, e.g. (R)- and (S)-mandelic acid. In particular, \textbf{3b} has the highest chiral selection efficiency with mandelate \cite{furusho_chirality-memory_1997, mizuno_chirality-memory_2000}. \textbf{1b} has no preference between the two enantiomers due to its high symmetry. Yamaguchi et al. resolved the x-ray crystal structure of the mandelate complex of \textbf{3b} and showed that the porphyrin adopts a saddle conformation, in which adjacent pyrrole units tilt alternatively \cite{mizuno_chirality-memory_2000}. This arrangement reduces the steric repulsion induced by aryl substitutions on the porphyrin periphery, resulting in equivalent up- and down-regions with respect to the quasi-molecular plane. In this complex, \textbf{3b} recognizes the chiral guest by two identical hydrogen bonds between \ce{N-H} bonds from two facing pyrroles and the carboxylate group, as the acidic proton is transferred completely to the porphyrin, along with additional non-covalent interactions with the porphyrin ring and/or substitutents. Such non-covalent interactions between the mandelate side chain and aryl groups on the porphyrin can play a critical role to favor one enantiomer over the other. Analyzing these nonspecific interactions is challenging, and thus there is no clear explanation towards the chiral recognition strength of these porphyrins. We propose that the unique chiral selection of the porphyrin is enabled by the existence of a local chiral environment generated by dissimilar substitutions at the \emph{meso} positions. The inner hydrogen-bonding region of both porphyrins \textbf{1b} and \textbf{3b} is highly symmetric (Figure~S3), leading to equivalent primary bonding interactions for both (R) and (S)-mandelic acid enantiomers. If the additional interactions with the mandelic acid side chain were also identical (or similar enough) for both enantiomers, then no chiral recognition would be possible. This situation could occur in two ways: first, there could be no chiral environment in the binding region which could generate a stereospecific response, and second, there could exist two local chiral environments, related by a global inversion or reflection, which could then bind the enantiomers equally. The latter case results in a globally achiral molecule such as \textbf{1b}. Thus, we propose that the existence of a strong, unique chiral environment at the substituent binding site is necessary for chiral recognition to occur.

First, we quantify the local chirality, which is equivalent to the chirotopicity concept of Mislow and Siegel \cite{mislow_stereoisomerism_1984}, at a grid of points around the porphyrins. The fractional lack of reflection or inversion symmetry centered at that point, for a sampling of the electronic density in the local region, therefore represents a chiral microenvironment localized at the evaluating point. 
%, indicated by negligible differences in the Wilcoxon statistical test across different porphyrins and radii (Figure \ref{fig:por_heatmap}).
Although only \textbf{1b} is achiral compared to the other chiral counterparts, the mean values of its chirotopicity field are consistently higher than for \textbf{2b}--\textbf{4b}. 

%\begin{figure}
%    \centering
%    \includegraphics[width=\linewidth]{heatmap.png}
%    \caption{Statistical difference test across different porphyrins and radii}
%    \label{fig:por_heatmap}
%\end{figure}

\begin{figure}
    \centering
    \includegraphics[width=\linewidth]{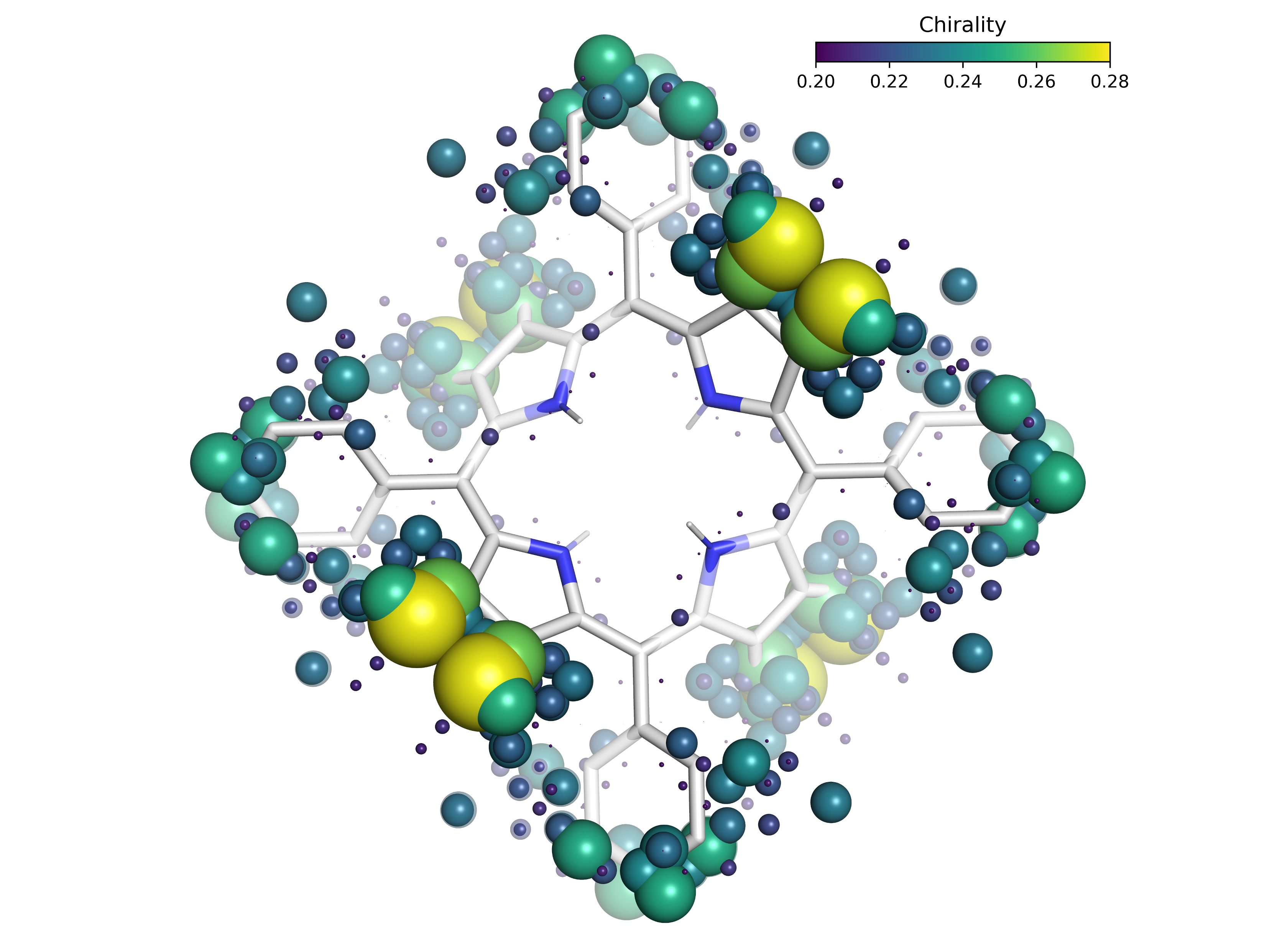}
    \caption{Chirotopicity field in unsubstituted porphyrin (\textbf{1b}). Larger, yellower spheres indicate a stronger local chiral environment.}
    \label{fig:por_base}
\end{figure}

%\begin{figure}
%    \centering
%    \includegraphics[width=\linewidth]{sub1_bar.png}
%    \caption{Chirotopicity field in monosubstituted porphyrin (2b) with (S)-mandelic acid (purple) and (R)-mandelic acid (orange)}
%    \label{fig:por_sub1}
%\end{figure}

\begin{figure}
    \centering
    \includegraphics[width=\linewidth]{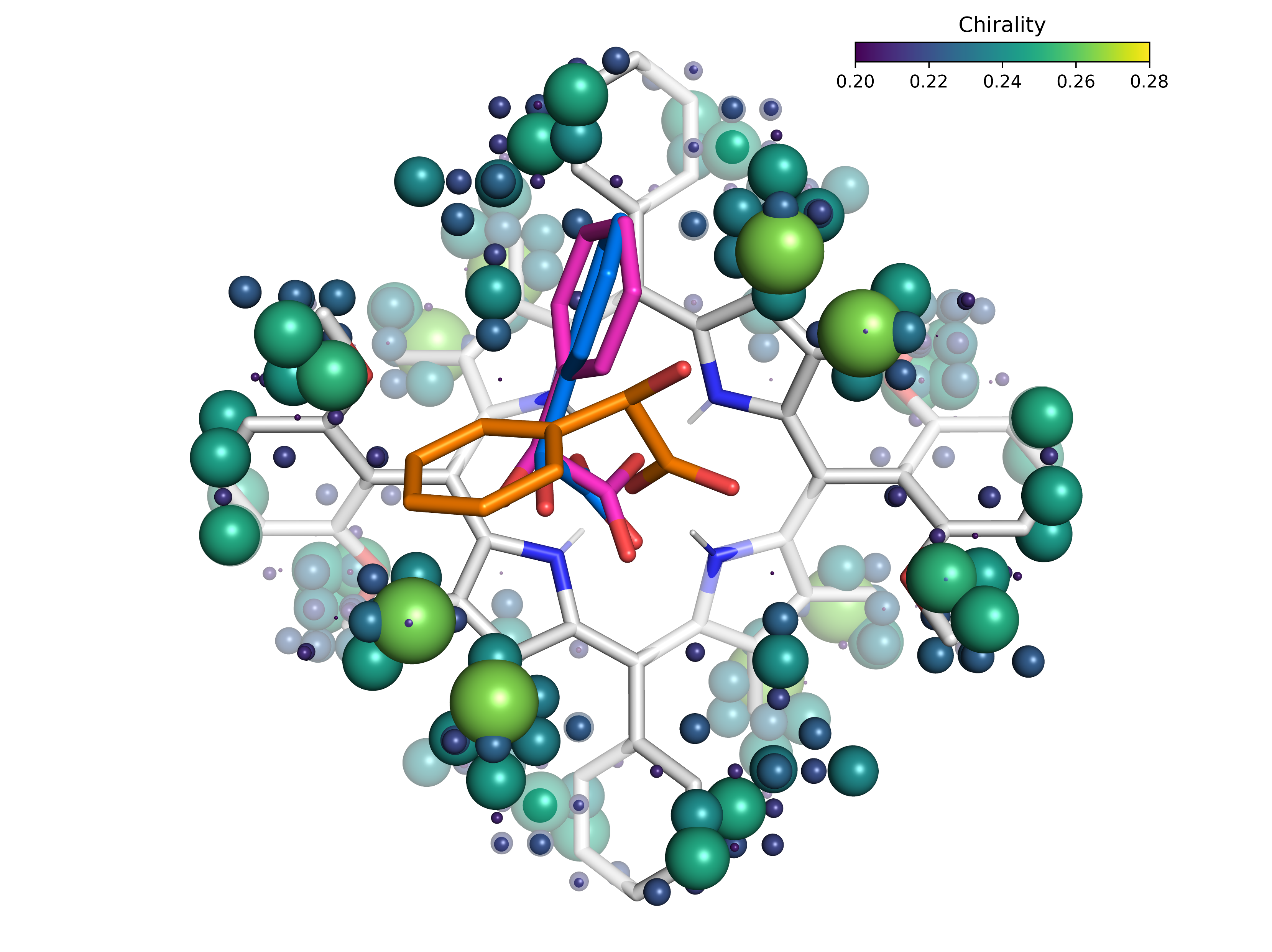}
    \caption{Chirotopicity field in disubstituted porphyrin (\textbf{3b}) with co-crytallized (S)-mandelic (blue), modeled (S)-mandelic acid (purple), and modeled (R)-mandelic acid (orange). Larger, yellower spheres indicate a stronger local chiral environment.}
    \label{fig:por_sub13}
\end{figure}

%\begin{figure}
%    \centering
%    \includegraphics[width=\linewidth]{sub123_bar.png}
%    \caption{Chirotopicity field in trisubstituted porphyrin (4b) with (S)-mandelic acid (purple) and (R)-mandelic acid (orange)}
    \label{fig:por_sub123}
%\end{figure}

Figures~\ref{fig:por_base}
%, \ref{fig:por_sub1},
and \ref{fig:por_sub13}
%, and \ref{fig:por_sub123}
illustrate the spatial distribution of the chirality field in porphyrins \textbf{1b} and \textbf{3b}, respectively. For the sake of visualization, we exclude points that have chirality lower than 0.2. High chirality is predominantly localized on peripheral substituents, while the inner porphyrin core remains largely achiral due to minimal electronic contributions from meso-substituents. The overall chiral patterns reflect the symmetry of the parent structure. In \textbf{1b}, the highest chirality regions are found around $\beta$-methyl groups, indicating a complex distortion involving the nitrogen atom and adjacent phenyl rings. The four \textit{meso}-phenyl groups present equivalent chiral microenvironments. Substituting a \textit{meso}-phenyl with a dimethoxyphenyl group in \textbf{2b} shifts the maximum chirality to the new substituent and neighboring phenyls (Figure~S5). However, the richer chirality on the two substitution-adjacent phenyl rings surprisingly points outward from the dimethoxyphenyl group, conserving the $C_2$ symmetry. The remaining phenyl shows reduced chirality, indicating a chirality transfer from C15 to C10 and C20 positions on the porphyrin core. With higher total symmetry, excessive chirality regions in \textbf{3b} (\Cref{fig:por_sub13}) reverts to the pyrrole which is similar to what observed in \textbf{1b}, suggesting that the molecular symmetry constrains the \textit{meso}-substituents, producing a more uniform chiral field. In \textbf{4b}, the highest chirality is associated with the dimethoxyphenyl group along the in-plane $C_2$ axis, but also localizes between the other two dimethoxyphenyl groups and on top of the pyrrole rings (Figure~S6). More importantly, there are high chirality corridors confined between an \emph{o}-dimethoxyphenyl, tilted-down pyrrole, and the phenyl groups, exhibited in chiral porphyrins \textbf{2b}--\textbf{4b}, but lacking in the achiral molecule \textbf{1b}. Due to the higher degree of symmetry, \textbf{3b} exhibits four chiral hostpots above each pyrrole, whereas \textbf{2b} and \textbf{4b} possess two such regions. 

%This chirality signature region is confirmed as recognition site between \textbf{3b} and mandelic acid guest. Moreover, the dominant number of chirality corridors on \textbf{3b} over \textbf{2b} and \textbf{4b} also justifies the greatest chirality-transfer efficiency of \textbf{3b} among the chiral porphyrins. 

We analyzed the interactions of (S)- and (R)-mandelic acid with chiral porphyrins \textbf{2b}--\textbf{4b} at the same level of theory \cite{marenich_universal_2009}. The porphyrin stereoisomers used here are favorable for binding to (S)-mandelate, and inverting the arrangement would favor the (R) enantiomer. As shown in \Cref{fig:por_sub13}, (S)-mandelate is stabilized above a region of relatively high chirotopicity. The shape of this region is also distinct, having no mirror- or inversion-symmetry analogue (c.f. the chiral regions above the pyrrole rings). Thus, we label this region as the chiral recognition site, as it permits a stereospecific binding interaction. On the other hand, the unfavorable (R)-mandelic acid binding configuration is oriented away from the recognition region, instead pointed toward the titled-up pyrrole. It is important to note that the putative (R) enantiomer binding site (no crystal structure was obtained in Ref.~\citenum{mizuno_chirality-memory_2000}) also experiences a relatively chiral local environment. \emph{There is no fundamental connection between the relative chirality of two binding regions and their relative binding strength.} Rather, local chirotopicity analysis must be combined with energy-based analyses such as non-covalent interaction (NCI) analysis \cite{johnsonRevealingNoncovalentInteractions2010}  or local energy decomposition \cite{altun_local_2021, bistoniFindingChemicalConcepts2020} in order to connect chirality to binding strength. In this case, the (R)-mandelic acid binding mode increases the steric repulsion with the dimethoxyphenyl group and destabilizes the complex. However, the strength of the local chiral environment in the chiral recognition site does suggest a potential qualitative connection with differential binding strength (at that site), as guest enantiomers will manifestly not experience the same intermolecular interactions. This effect is dependent on the ``overlap'' of the chiral and energetic features of the binding site, however. These findings demonstrate that the porphyrins' signature chiral chiral recognition site, a region whose combination of intermolecular interactions and chiral environment distinguishes \textbf{2b}--\textbf{4b} from \textbf{1b}, functions as a selective recognition factor for chiral guest molecules.

\section{Conclusion}

Although symmetry underlies a wide range of chemical concepts, current symmetry-driven theories have remained largely qualitative in explaining experimental phenomena. In this study, we introduce an algorithm that connects hidden local symmetry features with a continuous scale, enabling (semi-)quantitative correlations with molecular properties as well as numerical quantification of local chiral strength. Using local diene motifs in dendralenes as a case study, we demonstrated how their intrinsic symmetry governs molecular properties such as stability, Diels--Alder reactivity, and distinctive spectroscopic signatures. Furthermore, we presented a chirality field to quantify microscopic chirality embedded within a molecular framework. Notably, the chiral environment mapped for porphyrins reveals a distinct region that critically enables their enhanced chiral recognition toward chiral carboxylic acids.

Since the method captures both discrete and continuous considerations of symmetry at a fine-grained spatial resolution, it offers significant potential for broader applications. For example, one can apply the method to analyze dynamic symmetry breaking in reaction pathways, quantify conformational deformation and aggregation, and reveal non-specific interactions in complex molecular assemblies. Its generality also makes it suitable for studying symmetry-driven behavior in supramolecular systems, catalytic cavities, and bimolecular environments, where traditional symmetry descriptors exhibit limitations. The approach also opens new areas for systematically linking local symmetry metrics to structure-property relationships across various classes of chemical systems. Coupled with a local description of inter- and/or intra-molecular interaction energies, such an analysis also open the door for predictive calculations of chiral recognition and stereoselectivity phenomena.

\begin{acknowledgement}

This work was supported in part by the US National Science Foundation (grant CHE-2143725) and by the US Department of Energy (grant DE-SC0022893). Computational
resources for this research were provided by SMU’s O’Donnell Data Science and Research
Computing Institute.

\end{acknowledgement}

\begin{suppinfo}
Computational details and additional supplementary analysis; chirotopicity fields of \textbf{2b} and \textbf{4b}.
\end{suppinfo}

\bibliography{bib}

@article{petitjean_chirality_2003,
	title = {Chirality and {Symmetry} {Measures}: {A} {Transdisciplinary} {Review}},
	volume = {5},
	copyright = {http://creativecommons.org/licenses/by/3.0/},
	issn = {1099-4300},
	shorttitle = {Chirality and {Symmetry} {Measures}},
	url = {https://www.mdpi.com/1099-4300/5/3/271},
	doi = {10.3390/e5030271},
		number = {3},
	urldate = {2025-12-11},
	journal = {Entropy},
	author = {Petitjean, Michel},
	month = jul,
	year = {2003},
	note = {Company: Molecular Diversity Preservation International
Distributor: Molecular Diversity Preservation International
Institution: Molecular Diversity Preservation International
Label: Molecular Diversity Preservation International
Publisher: publisher},
	pages = {271--312},
}

@article{zabrodsky_continuous_1992,
	title = {Continuous symmetry measures},
	volume = {114},
	issn = {0002-7863},
	url = {https://doi.org/10.1021/ja00046a033},
	doi = {10.1021/ja00046a033},
	number = {20},
	urldate = {2025-12-11},
	journal = {Journal of the American Chemical Society},
	author = {Zabrodsky, Hagit and Peleg, Shmuel and Avnir, David},
	month = sep,
	year = {1992},
	note = {Publisher: American Chemical Society},
	pages = {7843--7851},
}

@article{woodward_conservation_1969,
	title = {The {Conservation} of {Orbital} {Symmetry}},
	volume = {8},
	copyright = {Copyright © 1969 by Verlag Chemie, GmbH, Germany},
	issn = {1521-3773},
	url = {https://onlinelibrary.wiley.com/doi/abs/10.1002/anie.196907811},
	doi = {10.1002/anie.196907811},
		number = {11},
	urldate = {2025-12-11},
	journal = {Angewandte Chemie International Edition in English},
	author = {Woodward, R. B. and Hoffmann, Roald},
	year = {1969},
	note = {\_eprint: https://onlinelibrary.wiley.com/doi/pdf/10.1002/anie.196907811},
	pages = {781--853},
}

@article{pearson_symmetry_1986,
	title = {Symmetry rules for chemical reactions},
	volume = {12},
	issn = {0898-1221},
	url = {https://www.sciencedirect.com/science/article/pii/0898122186901525},
	doi = {10.1016/0898-1221(86)90152-5},
	number = {1, Part B},
	urldate = {2025-12-11},
	journal = {Computers \& Mathematics with Applications},
	author = {Pearson, Ralph G.},
	month = jan,
	year = {1986},
	pages = {229--236},
}

@article{caldow_symmetry_1970,
	title = {Some symmetry aspects of homogeneous catalysis},
	volume = {6},
	issn = {0020-1650},
	url = {https://www.sciencedirect.com/science/article/pii/0020165070800642},
	doi = {10.1016/0020-1650(70)80064-2},
	number = {7},
	urldate = {2025-12-11},
	journal = {Inorganic and Nuclear Chemistry Letters},
	author = {Caldow, G. L. and MacGregor, R. A.},
	month = jul,
	year = {1970},
	pages = {645--649},
}

@article{maksic_symmetry_1986,
	title = {Symmetry, hybridization and bonding in molecules},
	volume = {12},
	issn = {0898-1221},
	url = {https://www.sciencedirect.com/science/article/pii/0898122186904190},
	doi = {10.1016/0898-1221(86)90419-0},
	number = {3, Part 2},
	urldate = {2025-12-11},
	journal = {Computers \& Mathematics with Applications},
	author = {Maksić, Zvonimir B.},
	month = may,
	year = {1986},
	pages = {697--723},
}

@article{hoffmann_interaction_1971,
	title = {Interaction of orbitals through space and through bonds},
	volume = {4},
	issn = {0001-4842},
	url = {https://doi.org/10.1021/ar50037a001},
	doi = {10.1021/ar50037a001},
	number = {1},
	urldate = {2025-12-11},
	journal = {Accounts of Chemical Research},
	author = {Hoffmann, Roald},
	month = jan,
	year = {1971},
	note = {Publisher: American Chemical Society},
	pages = {1--9},
}

@article{cram_host-guest_1974,
	title = {Host-{Guest} {Chemistry}},
	volume = {183},
	url = {https://www.science.org/doi/10.1126/science.183.4127.803},
	doi = {10.1126/science.183.4127.803},
	number = {4127},
	urldate = {2025-12-11},
	journal = {Science},
	author = {Cram, Donald J. and Cram, Jane M.},
	month = mar,
	year = {1974},
	note = {Publisher: American Association for the Advancement of Science},
	pages = {803--809},
}

@article{duarte_exploring_2022,
	title = {Exploring protein symmetry at the {RCSB} {Protein} {Data} {Bank}},
	volume = {6},
	issn = {2397-8554},
	url = {https://pmc.ncbi.nlm.nih.gov/articles/PMC9472815/},
	doi = {10.1042/ETLS20210267},
	number = {3},
	urldate = {2025-12-11},
	journal = {Emerging Topics in Life Sciences},
	author = {Duarte, Jose M. and Dutta, Shuchismita and Goodsell, David S. and Burley, Stephen K.},
	month = jul,
	year = {2022},
	pmid = {35801924},
	pmcid = {PMC9472815},
	pages = {231--243},
}

@article{mackay_demystifying_2013,
	title = {Demystifying the dendralenes},
	volume = {85},
	copyright = {De Gruyter expressly reserves the right to use all content for commercial text and data mining within the meaning of Section 44b of the German Copyright Act.},
	issn = {1365-3075},
	url = {https://www.degruyterbrill.com/document/doi/10.1351/PAC-CON-13-02-04/html},
	doi = {10.1351/PAC-CON-13-02-04},
		number = {6},
	urldate = {2025-12-12},
	journal = {Pure and Applied Chemistry},
	author = {Mackay, Emily G. and Sherburn, Michael S.},
	month = may,
	year = {2013},
	note = {Publisher: De Gruyter},
	pages = {1227--1239},
}

@article{hopf_dendralenesneglected_1984,
	title = {The {Dendralenes}—a {Neglected} {Group} of {Highly} {Unsaturated} {Hydrocarbons}},
	volume = {23},
	copyright = {Copyright © 1984 by Verlag Chemie, GmbH, Germany},
	issn = {1521-3773},
	url = {https://onlinelibrary.wiley.com/doi/abs/10.1002/anie.198409481},
	doi = {10.1002/anie.198409481},
		number = {12},
	urldate = {2025-12-12},
	journal = {Angewandte Chemie International Edition in English},
	author = {Hopf, Henning},
	year = {1984},
	note = {\_eprint: https://onlinelibrary.wiley.com/doi/pdf/10.1002/anie.198409481},
	pages = {948--960},
}

@article{george_general_2019,
	title = {A general synthesis of dendralenes},
	volume = {10},
	issn = {2041-6539},
	url = {https://pubs.rsc.org/en/content/articlelanding/2019/sc/c9sc03976g},
	doi = {10.1039/C9SC03976G},
		number = {43},
	urldate = {2025-12-12},
	journal = {Chemical Science},
	author = {George, Josemon and Ward, Jas S. and Sherburn, Michael S.},
	month = nov,
	year = {2019},
	note = {Publisher: The Royal Society of Chemistry},
	pages = {9969--9973},
}

@article{saglam_discovery_2016,
	title = {Discovery and {Computational} {Rationalization} of {Diminishing} {Alternation} in [n]{Dendralenes}},
	volume = {138},
	issn = {0002-7863},
	url = {https://doi.org/10.1021/jacs.5b11889},
	doi = {10.1021/jacs.5b11889},
	number = {3},
	urldate = {2025-12-12},
	journal = {Journal of the American Chemical Society},
	author = {Saglam, Mehmet F. and Fallon, Thomas and Paddon-Row, Michael N. and Sherburn, Michael S.},
	month = jan,
	year = {2016},
	note = {Publisher: American Chemical Society},
	pages = {1022--1032},
}

@article{mezey_degree_1991,
	title = {The degree of similarity of three-dimensional bodies: {Application} to molecular shape analysis},
	volume = {7},
	issn = {1572-8897},
	shorttitle = {The degree of similarity of three-dimensional bodies},
	url = {https://doi.org/10.1007/BF01200814},
	doi = {10.1007/BF01200814},
		number = {1},
	urldate = {2025-12-15},
	journal = {Journal of Mathematical Chemistry},
	author = {Mezey, Paul G.},
	month = dec,
	year = {1991},
	pages = {39--49},
}

@article{grimme_continuous_1998,
	title = {Continuous symmetry measures for electronic wavefunctions},
	volume = {297},
	issn = {0009-2614},
	url = {https://www.sciencedirect.com/science/article/pii/S0009261498011014},
	doi = {10.1016/S0009-2614(98)01101-4},
	number = {1},
	urldate = {2025-12-15},
	journal = {Chemical Physics Letters},
	author = {Grimme, Stefan},
	month = nov,
	year = {1998},
	pages = {15--22},
}

@article{guo_switchable_2022,
	title = {Switchable chiral transport in charge-ordered kagome metal {CsV3Sb5}},
	volume = {611},
	copyright = {2022 The Author(s)},
	issn = {1476-4687},
	url = {https://www.nature.com/articles/s41586-022-05127-9},
	doi = {10.1038/s41586-022-05127-9},
		number = {7936},
	urldate = {2025-12-15},
	journal = {Nature},
	author = {Guo, Chunyu and Putzke, Carsten and Konyzheva, Sofia and Huang, Xiangwei and Gutierrez-Amigo, Martin and Errea, Ion and Chen, Dong and Vergniory, Maia G. and Felser, Claudia and Fischer, Mark H. and Neupert, Titus and Moll, Philip J. W.},
	month = nov,
	year = {2022},
	note = {Publisher: Nature Publishing Group},
	pages = {461--466},
}

@article{shah_calochorturils_2025,
	title = {Calochorturils: {Chiral} {Bowl}-{Shaped} {Cavitands} {Obtained} by {Anisotropic} {Tangential} {Substitution}},
	volume = {147},
	issn = {0002-7863},
	shorttitle = {Calochorturils},
	url = {https://doi.org/10.1021/jacs.5c11855},
	doi = {10.1021/jacs.5c11855},
	number = {42},
	urldate = {2025-12-15},
	journal = {Journal of the American Chemical Society},
	author = {Shah, Sadhna and Ganga, Venkata S. R. and Huang, Yu-Xiang and Fridman, Natalia and Tuvi-Arad, Inbal and Chan, Yi-Tsu and Reany, Ofer and Keinan, Ehud},
	month = oct,
	year = {2025},
	note = {Publisher: American Chemical Society},
	pages = {38443--38451},
}

@article{grieder_relation_2025,
	title = {Relation of {Continuous} {Chirality} {Measure} to {Spin} and {Orbital} {Polarization}, and {Chiroptical} {Properties} in {Solids}},
	volume = {13},
	copyright = {© 2025 The Author(s). Advanced Optical Materials published by Wiley-VCH GmbH},
	issn = {2195-1071},
	url = {https://onlinelibrary.wiley.com/doi/abs/10.1002/adom.202501190},
	doi = {10.1002/adom.202501190},
		number = {33},
	urldate = {2025-12-15},
	journal = {Advanced Optical Materials},
	author = {Grieder, Andrew and Tu, Shihao and Ping, Yuan},
	year = {2025},
	note = {\_eprint: https://advanced.onlinelibrary.wiley.com/doi/pdf/10.1002/adom.202501190},
	pages = {e01190},
}

@article{mislow_stereoisomerism_1984,
	title = {Stereoisomerism and local chirality},
	volume = {106},
	issn = {0002-7863},
	url = {https://doi.org/10.1021/ja00323a043},
	doi = {10.1021/ja00323a043},
	number = {11},
	urldate = {2025-12-15},
	journal = {Journal of the American Chemical Society},
	author = {Mislow, Kurt and Siegel, Jay},
	month = may,
	year = {1984},
	note = {Publisher: American Chemical Society},
	pages = {3319--3328},
}

@article{lipinski_local_2014,
	title = {Local chirality measures in {QSPR} : {IR} and {VCD} spectroscopy},
	volume = {4},
	issn = {2046-2069},
	shorttitle = {Local chirality measures in {QSPR}},
	url = {https://pubs.rsc.org/en/content/articlelanding/2014/ra/c4ra08434a},
	doi = {10.1039/C4RA08434A},
		number = {87},
	urldate = {2025-12-15},
	journal = {RSC Advances},
	author = {Lipiński, Piotr F. J. and Dobrowolski, Jan Cz},
	month = sep,
	year = {2014},
	note = {Publisher: The Royal Society of Chemistry},
	pages = {47047--47055},
}

@article{alvarez_continuous_2005,
	title = {Continuous chirality measures in transition metal chemistry},
	volume = {34},
	issn = {1460-4744},
	url = {https://pubs.rsc.org/en/content/articlelanding/2005/cs/b301406c},
	doi = {10.1039/B301406C},
		number = {4},
	urldate = {2025-12-15},
	journal = {Chemical Society Reviews},
	author = {Alvarez, Santiago and Alemany, Pere and Avnir, David},
	month = mar,
	year = {2005},
	note = {Publisher: The Royal Society of Chemistry},
	pages = {313--326},
}

@article{payne_practical_2009,
	title = {Practical {Synthesis} of the {Dendralene} {Family} {Reveals} {Alternation} in {Behavior}},
	volume = {48},
	copyright = {Copyright © 2009 WILEY-VCH Verlag GmbH \& Co. KGaA, Weinheim},
	issn = {1521-3773},
	url = {https://onlinelibrary.wiley.com/doi/abs/10.1002/anie.200901733},
	doi = {10.1002/anie.200901733},
		number = {26},
	urldate = {2025-12-15},
	journal = {Angewandte Chemie International Edition},
	author = {Payne, Alan D. and Bojase, Gomotsang and Paddon-Row, Michael N. and Sherburn, Michael S.},
	year = {2009},
	note = {\_eprint: https://onlinelibrary.wiley.com/doi/pdf/10.1002/anie.200901733},
	pages = {4836--4839},
}

@article{saglam_synthesis_2016,
	title = {Synthesis and {Diels}–{Alder} {Reactivity} of {Substituted} [4]{Dendralenes}},
	volume = {81},
	issn = {0022-3263},
	url = {https://doi.org/10.1021/acs.joc.5b02583},
	doi = {10.1021/acs.joc.5b02583},
	number = {4},
	urldate = {2025-12-15},
	journal = {The Journal of Organic Chemistry},
	author = {Saglam, Mehmet F. and Alborzi, Ali R. and Payne, Alan D. and Willis, Anthony C. and Paddon-Row, Michael N. and Sherburn, Michael S.},
	month = feb,
	year = {2016},
	note = {Publisher: American Chemical Society},
	pages = {1461--1475},
}

@article{pronin_synthesis_2012,
	title = {Synthesis of a {Potent} {Antimalarial} {Amphilectene}},
	volume = {134},
	issn = {0002-7863},
	url = {https://doi.org/10.1021/ja310129b},
	doi = {10.1021/ja310129b},
	number = {48},
	urldate = {2025-12-15},
	journal = {Journal of the American Chemical Society},
	author = {Pronin, Sergey V. and Shenvi, Ryan A.},
	month = dec,
	year = {2012},
	note = {Publisher: American Chemical Society},
	pages = {19604--19606},
}

@article{fielder_first_2000,
	title = {First {Synthesis} of the {Dendralene} {Family} of {Fundamental} {Hydrocarbons}},
	volume = {39},
	copyright = {Copyright © 2000 WILEY-VCH Verlag GmbH, Weinheim, Fed. Rep. of Germany},
	issn = {1521-3773},
	url = {https://onlinelibrary.wiley.com/doi/abs/10.1002/1521-3773%2820001201%2939%3A23%3C4331%3A%3AAID-ANIE4331%3E3.0.CO%3B2-3},
	doi = {10.1002/1521-3773(20001201)39:23<4331::AID-ANIE4331>3.0.CO;2-3},
	number = {23},
	urldate = {2025-12-15},
	journal = {Angewandte Chemie International Edition},
	author = {Fielder, Simon and Rowan, Daryl D. and Sherburn, Michael S.},
	year = {2000},
	note = {\_eprint: https://onlinelibrary.wiley.com/doi/pdf/10.1002/1521-3773\%2820001201\%2939\%3A23\%3C4331\%3A\%3AAID-ANIE4331\%3E3.0.CO\%3B2-3},
	pages = {4331--4333},
}

@article{woo_indium-mediated_1999,
	title = {Indium-{Mediated} $\gamma$-{Pentadienylation} of {Aldehydes} and {Ketones}: {Cross}-{Conjugated} {Trienes} for {Diene}-{Transmissive} {Cycloadditions}},
	volume = {1},
	issn = {1523-7060},
	shorttitle = {Indium-{Mediated} $\gamma$-{Pentadienylation} of {Aldehydes} and {Ketones}},
	url = {https://doi.org/10.1021/ol990695c},
	doi = {10.1021/ol990695c},
	number = {4},
	urldate = {2025-12-15},
	journal = {Organic Letters},
	author = {Woo, Simon and Squires, Neil and Fallis, Alex G.},
	month = aug,
	year = {1999},
	note = {Publisher: American Chemical Society},
	pages = {573--576},
}

@article{syrbu_phenyl-substituted_2004,
	title = {Phenyl-substituted porphyrins. 1. {Synthesis} of meso-phenyl-substituted porphyrins},
	volume = {40},
	issn = {1573-8353},
	url = {https://doi.org/10.1007/s10593-005-0050-6},
	doi = {10.1007/s10593-005-0050-6},
	number = {10},
	urldate = {2026-02-19},
	journal = {Chem Heterocycl Compd},
	author = {Syrbu, S. A. and Lyubimova, T. V. and Semeikin, A. S.},
	month = oct,
	year = {2004},
	pages = {1262--1270},
}

@article{keinan_quantitative_2000,
	title = {Quantitative {Symmetry} in {Structure}--{Activity} {Correlations}: {The} {Near} {C2} {Symmetry} of {Inhibitor}/{HIV} {Protease} {Complexes}},
	volume = {122},
	issn = {0002-7863},
	shorttitle = {Quantitative {Symmetry} in {Structure}--{Activity} {Correlations}},
	url = {https://doi.org/10.1021/ja993810n},
	doi = {10.1021/ja993810n},
	number = {18},
	urldate = {2025-12-15},
	journal = {Journal of the American Chemical Society},
	author = {Keinan, Shahar and Avnir, David},
	month = may,
	year = {2000},
	note = {Publisher: American Chemical Society},
	pages = {4378--4384},
}

@article{marenich_universal_2009,
	title = {Universal {Solvation} {Model} {Based} on {Solute} {Electron} {Density} and on a {Continuum} {Model} of the {Solvent} {Defined} by the {Bulk} {Dielectric} {Constant} and {Atomic} {Surface} {Tensions}},
	volume = {113},
	issn = {1520-6106},
	url = {https://doi.org/10.1021/jp810292n},
	doi = {10.1021/jp810292n},
	number = {18},
	urldate = {2025-12-15},
	journal = {The Journal of Physical Chemistry B},
	author = {Marenich, Aleksandr V. and Cramer, Christopher J. and Truhlar, Donald G.},
	month = may,
	year = {2009},
	note = {Publisher: American Chemical Society},
	pages = {6378--6396},
}

@article{sang_symmetry_2019,
	title = {Symmetry {Breaking} in {Self}-{Assembled} {Nanoassemblies}},
	volume = {11},
	copyright = {http://creativecommons.org/licenses/by/3.0/},
	issn = {2073-8994},
	url = {https://www.mdpi.com/2073-8994/11/8/950},
	doi = {10.3390/sym11080950},
		number = {8},
	urldate = {2025-12-15},
	journal = {Symmetry},
	author = {Sang, Yutao and Liu, Minghua and Sang, Yutao and Liu, Minghua},
	month = jul,
	year = {2019},
	note = {Company: Multidisciplinary Digital Publishing Institute
Distributor: Multidisciplinary Digital Publishing Institute
Institution: Multidisciplinary Digital Publishing Institute
Label: Multidisciplinary Digital Publishing Institute
Publisher: publisher},
}

@article{sang_hierarchical_2022,
	title = {Hierarchical self-assembly into chiral nanostructures},
	volume = {13},
	url = {https://pubs.rsc.org/en/content/articlelanding/2022/sc/d1sc03561d},
	doi = {10.1039/D1SC03561D},
		number = {3},
	urldate = {2025-12-15},
	journal = {Chemical Science},
	author = {Sang, Yutao and Liu, Minghua},
	year = {2022},
	note = {Publisher: Royal Society of Chemistry},
	pages = {633--656},
}

@article{shi_symmetry_2016,
	title = {Symmetry breaking in molecular ferroelectrics},
	volume = {45},
	issn = {1460-4744},
	url = {https://pubs.rsc.org/en/content/articlelanding/2016/cs/c5cs00308c},
	doi = {10.1039/C5CS00308C},
		number = {14},
	urldate = {2025-12-15},
	journal = {Chemical Society Reviews},
	author = {Shi, Ping-Ping and Tang, Yuan-Yuan and Li, Peng-Fei and Liao, Wei-Qiang and Wang, Zhong-Xia and Ye, Qiong and Xiong, Ren-Gen},
	month = jul,
	year = {2016},
	note = {Publisher: The Royal Society of Chemistry},
	pages = {3811--3827},
}

@article{bersuker_jahn-teller_2017,
	title = {The {Jahn}-{Teller} and pseudo {Jahn}-{Teller} effect in materials science},
	volume = {833},
	issn = {1742-6596},
	url = {https://doi.org/10.1088/1742-6596/833/1/012001},
	doi = {10.1088/1742-6596/833/1/012001},
		number = {1},
	urldate = {2025-12-15},
	journal = {Journal of Physics: Conference Series},
	author = {Bersuker, I B},
	month = apr,
	year = {2017},
	note = {Publisher: IOP Publishing},
	pages = {012001},
}

@article{bersuker_jahnteller_2021,
	title = {Jahn–{Teller} and {Pseudo}-{Jahn}–{Teller} {Effects}: {From} {Particular} {Features} to {General} {Tools} in {Exploring} {Molecular} and {Solid} {State} {Properties}},
	volume = {121},
	issn = {0009-2665},
	shorttitle = {Jahn–{Teller} and {Pseudo}-{Jahn}–{Teller} {Effects}},
	url = {https://doi.org/10.1021/acs.chemrev.0c00718},
	doi = {10.1021/acs.chemrev.0c00718},
	number = {3},
	urldate = {2025-12-15},
	journal = {Chemical Reviews},
	author = {Bersuker, Isaac B.},
	month = feb,
	year = {2021},
	note = {Publisher: American Chemical Society},
	pages = {1463--1512},
}

@article{aucar_relationship_2024,
	title = {A {Relationship} between the {Molecular} {Parity}-{Violation} {Energy} and the {Electronic} {Chirality} {Measure}},
	volume = {15},
	url = {https://doi.org/10.1021/acs.jpclett.3c03038},
	doi = {10.1021/acs.jpclett.3c03038},
	number = {1},
	urldate = {2025-12-15},
	journal = {The Journal of Physical Chemistry Letters},
	author = {Aucar, Juan J. and Stroppa, Alessandro and Aucar, Gustavo A.},
	month = jan,
	year = {2024},
	note = {Publisher: American Chemical Society},
	pages = {234--240},
}

@article{ferrarini_assessment_1998,
	title = {On the assessment of molecular chirality},
	issn = {1364-5471},
	url = {https://pubs.rsc.org/en/content/articlelanding/1998/p2/a702619f},
	doi = {10.1039/A702619F},
		number = {2},
	urldate = {2025-12-15},
	journal = {Journal of the Chemical Society, Perkin Transactions 2},
	author = {Ferrarini, Alberta and Nordio, Pier Luigi},
	month = jan,
	year = {1998},
	note = {Publisher: The Royal Society of Chemistry},
	pages = {455--460},
}

@article{mizuno_chirality-memory_2000,
	title = {Chirality-{Memory} {Molecule}: {Crystallographic} and {Spectroscopic} {Studies} on {Dynamic} {Molecular} {Recognition} {Events} by {Fully} {Substituted} {Chiral} {Porphyrins}},
	volume = {122},
	issn = {0002-7863},
	shorttitle = {Chirality-{Memory} {Molecule}},
	url = {https://doi.org/10.1021/ja000052o},
	doi = {10.1021/ja000052o},
	number = {22},
	urldate = {2025-12-15},
	journal = {Journal of the American Chemical Society},
	author = {Mizuno, Yukitami and Aida, Takuzo and Yamaguchi, Kentaro},
	month = jun,
	year = {2000},
	note = {Publisher: American Chemical Society},
	pages = {5278--5285},
}

@article{furusho_chirality-memory_1997,
	title = {Chirality-{Memory} {Molecule}: {A} {D2}-{Symmetric} {Fully} {Substituted} {Porphyrin} as a {Conceptually} {New} {Chirality} {Sensor}},
	volume = {119},
	issn = {0002-7863},
	shorttitle = {Chirality-{Memory} {Molecule}},
	url = {https://doi.org/10.1021/ja970431q},
	doi = {10.1021/ja970431q},
	number = {22},
	urldate = {2025-12-15},
	journal = {Journal of the American Chemical Society},
	author = {Furusho, Yoshio and Kimura, Takayuki and Mizuno, Yukitami and Aida, Takuzo},
	month = jun,
	year = {1997},
	note = {Publisher: American Chemical Society},
	pages = {5267--5268},
}

@article{fowler_vocabulary_1992,
	title = {Vocabulary for fuzzy symmetry},
	volume = {360},
	copyright = {1992 Springer Nature Limited},
	issn = {1476-4687},
	url = {https://www.nature.com/articles/360626a0},
	doi = {10.1038/360626a0},
		number = {6405},
	urldate = {2025-12-19},
	journal = {Nature},
	author = {Fowler, Patrick W.},
	month = dec,
	year = {1992},
	note = {Publisher: Nature Publishing Group},
	pages = {626--626},
}

@article{hirotoSynthesisFunctionalizationPorphyrins2017,
    title = {Synthesis and {Functionalization} of {Porphyrins} through {Organometallic} {Methodologies}},
    volume = {117},
    issn = {0009-2665},
    url = {https://doi.org/10.1021/acs.chemrev.6b00427},
    doi = {10.1021/acs.chemrev.6b00427},
    number = {4},
    urldate = {2026-01-15},
    journal = {Chemical Reviews},
    publisher = {American Chemical Society},
    author = {Hiroto, Satoru and Miyake, Yoshihiro and Shinokubo, Hiroshi},
    month = feb,
    year = {2017},
    pages = {2910--3043},
}

@article{toombs-ruane_dielsalder_2012,
    title = {On the {Diels}–{Alder} dimerisation of cross-conjugated trienes},
    volume = {48},
    issn = {1364-548X},
    url = {https://pubs.rsc.org/en/content/articlelanding/2012/cc/c2cc32520a},
    doi = {10.1039/C2CC32520A},
    number = {53},
    urldate = {2026-03-07},
    journal = {Chemical Communications},
    publisher = {The Royal Society of Chemistry},
    author = {Toombs-Ruane, Henry and Pearson, Emma L. and Paddon-Row, Michael N. and Sherburn, Michael S.},
    month = jun,
    year = {2012},
    pages = {6639--6641},
}

@article{paddon-row_origin_2011,
    title = {On the origin of the alternating {Diels}–{Alder} reactivity in [n]dendralenes},
    volume = {48},
    issn = {1364-548X},
    url = {https://pubs.rsc.org/en/content/articlelanding/2012/cc/c1cc15455a},
    doi = {10.1039/C1CC15455A},
    number = {6},
    urldate = {2026-03-07},
    journal = {Chemical Communications},
    publisher = {The Royal Society of Chemistry},
    author = {Paddon-Row, Michael N. and Sherburn, Michael S.},
    month = dec,
    year = {2011},
    pages = {832--834},
}

@article{cui_bionic_2024,
    title = {Bionic nanopore recognition receptors for single-molecule enantioselectivity studies of chiral drugs},
    volume = {1318},
    issn = {0003-2670},
    url = {https://www.sciencedirect.com/science/article/pii/S000326702400761X},
    doi = {10.1016/j.aca.2024.342960},
    urldate = {2026-03-07},
    journal = {Analytica Chimica Acta},
    author = {Cui, Rikun and Wang, Zhenzhao and Li, Linna and Liu, Lili and Li, Zhen and Liu, Xingtong and Chen, Tingting and Rauf, Ayesha and Kang, Xiaofeng and Guo, Yanli},
    month = aug,
    year = {2024},
    pages = {342960},
}

@article{daskova_turning_2022,
    title = {Turning {Enantiomeric} {Relationships} into {Diastereomeric} {Ones}: {Self}-{Resolving} $\alpha$-{Ureidophosphonates} and {Their} {Organocatalytic} {Enantioselective} {Synthesis}},
    volume = {144},
    issn = {0002-7863},
    shorttitle = {Turning {Enantiomeric} {Relationships} into {Diastereomeric} {Ones}},
    url = {https://doi.org/10.1021/jacs.2c10911},
    doi = {10.1021/jacs.2c10911},
    number = {51},
    urldate = {2026-03-07},
    journal = {Journal of the American Chemical Society},
    publisher = {American Chemical Society},
    author = {Dašková, Vanda and Padín, Damián and Feringa, Ben L.},
    month = dec,
    year = {2022},
    pages = {23603--23613},
}

@article{che_hierarchical_2025,
    title = {Hierarchical {Porous} {Homochiral} {Metal}-{Organic} {Frameworks} for {Enantiomeric} {Separation} of {Chiral} {Drugs}},
    volume = {35},
    issn = {1616-301X},
    url = {https://doi.org/10.1002/adfm.202506753},
    doi = {10.1002/adfm.202506753},
    number = {36},
    urldate = {2026-03-07},
    journal = {Advanced Functional Materials},
    publisher = {John Wiley \& Sons, Ltd},
    author = {Che, Yuanyuan and Li, Ke and Xie, Cheng and Sun, Fuyi and Yuan, Yan and Zhao, Xunke and Zhao, Chun-Xia and Chang, Ganggang},
    month = sep,
    year = {2025},
    pages = {2506753},
}

@book{berthod_chiral_2010,
    address = {Berlin, Heidelberg},
    title = {Chiral {Recognition} in {Separation} {Methods}},
    copyright = {http://www.springer.com/tdm},
    isbn = {978-3-642-12444-0 978-3-642-12445-7},
    url = {http://link.springer.com/10.1007/978-3-642-12445-7},
    doi = {10.1007/978-3-642-12445-7},
    urldate = {2026-03-07},
    publisher = {Springer},
    editor = {Berthod, Alain},
    year = {2010},
}

@article{mayer_nonlinear_2022,
    title = {Nonlinear {Effects} in {Asymmetric} {Catalysis} by {Design}: {Concept}, {Synthesis}, and {Applications}},
    volume = {55},
    issn = {0001-4842},
    shorttitle = {Nonlinear {Effects} in {Asymmetric} {Catalysis} by {Design}},
    url = {https://doi.org/10.1021/acs.accounts.2c00557},
    doi = {10.1021/acs.accounts.2c00557},
    number = {23},
    urldate = {2026-03-07},
    journal = {Accounts of Chemical Research},
    publisher = {American Chemical Society},
    author = {Mayer, Lena C. and Heitsch, Simone and Trapp, Oliver},
    month = dec,
    year = {2022},
    pages = {3345--3361},
}

@article{sallembien_possible_2022,
    title = {Possible chemical and physical scenarios towards biological homochirality},
    volume = {51},
    issn = {1460-4744},
    url = {https://pubs.rsc.org/en/content/articlelanding/2022/cs/d1cs01179k},
    doi = {10.1039/D1CS01179K},
    number = {9},
    urldate = {2026-03-07},
    journal = {Chemical Society Reviews},
    publisher = {The Royal Society of Chemistry},
    author = {Sallembien, Quentin and Bouteiller, Laurent and Crassous, Jeanne and Raynal, Matthieu},
    month = may,
    year = {2022},
    pages = {3436--3476},
}

@article{pavlov_hiral_2019,
    title = {Chiral and {Racemic} {Fields} {Concept} for {Understanding} of the {Homochirality} {Origin}, {Asymmetric} {Catalysis}, {Chiral} {Superstructure} {Formation} from {Achiral} {Molecules}, and {B}-{Z} {DNA} {Conformational} {Transition}},
    volume = {11},
    copyright = {http://creativecommons.org/licenses/by/3.0/},
    issn = {2073-8994},
    url = {https://www.mdpi.com/2073-8994/11/5/649},
    doi = {10.3390/sym11050649},
    number = {5},
    urldate = {2026-03-07},
    journal = {Symmetry},
    publisher = {Multidisciplinary Digital Publishing Institute},
    author = {Pavlov, Valerii A. and Shushenachev, Yaroslav V. and Zlotin, Sergey G.},
    month = may,
    year = {2019},
    pages = {649},
}

@article{takahashi_origin_2019,
    title = {Origin of {Terrestrial} {Bioorganic} {Homochirality} and {Symmetry} {Breaking} in the {Universe}},
    volume = {11},
    copyright = {http://creativecommons.org/licenses/by/3.0/},
    issn = {2073-8994},
    url = {https://www.mdpi.com/2073-8994/11/7/919},
    doi = {10.3390/sym11070919},
    number = {7},
    urldate = {2026-03-07},
    journal = {Symmetry},
    publisher = {Multidisciplinary Digital Publishing Institute},
    author = {Takahashi, Jun-ichi and Kobayashi, Kensei},
    month = jul,
    year = {2019},
    pages = {919},
}

@article{buhse_spontaneous_2021,
    title = {Spontaneous {Deracemizations}},
    volume = {121},
    issn = {0009-2665},
    url = {https://doi.org/10.1021/acs.chemrev.0c00819},
    doi = {10.1021/acs.chemrev.0c00819},
    number = {4},
    urldate = {2026-03-07},
    journal = {Chemical Reviews},
    publisher = {American Chemical Society},
    author = {Buhse, Thomas and Cruz, José-Manuel and Noble-Terán, María E. and Hochberg, David and Ribó, Josep M. and Crusats, Joaquim and Micheau, Jean-Claude},
    month = feb,
    year = {2021},
    pages = {2147--2229},
}

@article{nguyen_chiral_2006,
    title = {Chiral {Drugs}: {An} {Overview}},
    volume = {2},
    issn = {1550-9702},
    shorttitle = {Chiral {Drugs}},
    url = {https://pmc.ncbi.nlm.nih.gov/articles/PMC3614593/},
    number = {2},
    urldate = {2026-03-07},
    journal = {International Journal of Biomedical Science : IJBS},
    author = {Nguyen, Lien Ai and He, Hua and Pham-Huy, Chuong},
    month = jun,
    year = {2006},
    pages = {85--100},
}

@article{altun_local_2021,
    title = {Local energy decomposition of coupled-cluster interaction energies: {Interpretation}, benchmarks, and comparison with symmetry-adapted perturbation theory},
    volume = {121},
    copyright = {© 2020 The Authors. International Journal of Quantum Chemistry published by Wiley Periodicals LLC.},
    issn = {1097-461X},
    shorttitle = {Local energy decomposition of coupled-cluster interaction energies},
    url = {https://onlinelibrary.wiley.com/doi/abs/10.1002/qua.26339},
    doi = {10.1002/qua.26339},
    number = {3},
    urldate = {2026-03-13},
    journal = {International Journal of Quantum Chemistry},
    author = {Altun, Ahmet and Izsák, Róbert and Bistoni, Giovanni},
    year = {2021},
    note = {\_eprint: https://onlinelibrary.wiley.com/doi/pdf/10.1002/qua.26339},
    pages = {e26339},
}

@article{bistoniFindingChemicalConcepts2020,
    title = {Finding chemical concepts in the {Hilbert} space: {Coupled} cluster analyses of noncovalent interactions},
    volume = {10},
    copyright = {© 2019 The Author. WIREs Computational Molecular Science published by Wiley Periodicals, Inc.},
    issn = {1759-0884},
    shorttitle = {Finding chemical concepts in the {Hilbert} space},
    url = {https://onlinelibrary.wiley.com/doi/abs/10.1002/wcms.1442},
    doi = {10.1002/wcms.1442},
    number = {3},
    urldate = {2026-03-13},
    journal = {WIREs Computational Molecular Science},
    author = {Bistoni, Giovanni},
    year = {2020},
    note = {\_eprint: https://wires.onlinelibrary.wiley.com/doi/pdf/10.1002/wcms.1442},
    pages = {e1442},
}

@article{johnsonRevealingNoncovalentInteractions2010,
    title = {Revealing {Noncovalent} {Interactions}},
    volume = {132},
    issn = {0002-7863},
    url = {https://doi.org/10.1021/ja100936w},
    doi = {10.1021/ja100936w},
    number = {18},
    urldate = {2026-03-13},
    journal = {Journal of the American Chemical Society},
    publisher = {American Chemical Society},
    author = {Johnson, Erin R. and Keinan, Shahar and Mori-Sánchez, Paula and Contreras-García, Julia and Cohen, Aron J. and Yang, Weitao},
    month = may,
    year = {2010},
    pages = {6498--6506},
}

\end{document}